\documentclass[10pt]{article}

\usepackage[letterpaper]{geometry}
\usepackage{hicss51}
\usepackage{times}
\usepackage[none]{hyphenat}
\usepackage[hyphens]{url}
\usepackage{latexsym}
\usepackage[frozencache=true]{minted}
\usepackage{indentfirst}
\usepackage{graphicx}
\graphicspath{{images/}}

\usepackage{subcaption}
\usepackage[table]{xcolor}

\usepackage{hyperref}
\usepackage{doi}
\hypersetup{
    colorlinks=true,
    linkcolor=blue,
    filecolor=magenta,      
    urlcolor=cyan,
}

\title{Media Accessibility Policy in Theory and Reality: \\ Empirical Outreach to Audio Description Users in the United States\thanks{*This paper has been accepted for the upcoming 52\textsuperscript{nd} Hawaii International Conference on System Sciences (HICSS-52). Publication in the Conference Proceedings is pending on author's presentation of this paper at the conference.}}

 \author{Philipp Jordan \\ Communication and Information Sciences \\ University of Hawai`i at M{\=a}noa \\
 {\href{mailto:philippj@hawaii.edu}{philippj@hawaii.edu}}\\ \And
   Brett Oppegaard \\ School of Communications \\ University of Hawai`i at M{\=a}noa \\
   {\href{mailto:brett.oppegaard@hawaii.edu}{brett.oppegaard@hawaii.edu}} \\
 }

\date{12.09.2018}

\begin{document}
\maketitle
\begin{abstract} 

Audio description, a form of trans-modal media translation, allows people who are blind or visually impaired access to visually-oriented, socio-cultural, or historical public discourse alike. Although audio description has gained more prominence in media policy and research lately, it rarely has been studied empirically. Yet this paper presents quantitative and qualitative survey data on its challenges and opportunities, through the analysis of responses from 483 participants in a national sample, with 334 of these respondents being blind. Our results give insight into audio description use in broadcast TV, streaming services, for physical media, such as DVDs, and in movie theaters. We further discover a multiplicity of barriers and hindrances which prevent a better adoption and larger proliferation of audio description. In our discussion, we present a possible answer to these problems -- the UniDescription Project -- a media ecosystem for the creation, curation, and dissemination of audio description for multiple media platforms.
\end{abstract}

\section{Introduction}For visually impaired or blind audiences, audio description (AD) \cite{Braun2008,Braun2011,Matamala2016,Fryer.2016} is an effective way to provide equivalent access and opportunity for participation in social discourse, including through art and cultural exhibits, across educational and professional communities. Since its inception in the 1970s, the field has grown considerably. In the last decades, a variety of policies, regulations, and laws have been passed and implemented in the United States (US), and elsewhere, to require the provision of an alternative format of entertainment in broadcast Television (TV), streaming services and the cinema industry. 

Due to these policies -- for example the 21\textsuperscript{st} Century Communications and Video Accessibility Act (CCVAA) \cite{FCC2018} -- AD  has become more widely available in TV and movie theaters, as well as streaming services \cite{NetflixAD}. 

\noindent While we  can observe and measure a gradual growth of audio-described material across a spectrum of broadcast-, streaming- and screening services (e.g. \cite{ADP_AD_master}), we find little (if any) empirical efforts to encapsulate AD and its impacts, including accounts of primary target audiences perceptions, plus their opinions and receptions of these offerings. Through a partnership with the American Council of the Blind (ACB), who commissioned this national survey, but has not shared its detailed data elsewhere, this study will report on a novel and unique dataset of the uses, challenges, and limitations of AD, assessed by its primary intended users, who are visually impaired or blind.

\section{Background}
Policy implementations to provide equal access to people with cognitive or physical disabilities have been progressing in the last three  decades in the US. From the principles and policies in Sections 504 \cite{Rehab_act_1973_504} and 508 \cite{Rehab_act_1973_508} of the Rehabilitation Act of 1973, the passing of the Americans with Disabilities Act in 1990 \cite{Disability_Act1990} and recently, Title II of the 21\textsuperscript{st} CCVAA \cite{FCC2018} in 2010 with a focus on equal access to traditional media and mobile technologies. 

It is important to note that these laws and policies are continuously adapted, amended and modified by for instance, the Federal Communications Commission (FCC), which can equally grant waivers for specific media providers and hear petitions from accessibility advocacy groups alike. For example, the June 2017 FCC Factsheet on Video Description Expansion on the grounds of the 21\textsuperscript{st} CCVAA \cite[p.5]{FCC_Fact_Sheet_2017} envisioned that: 
\begin{quote}
\small
\textit{``Ideally, viewers who are blind or visually impaired would have the same range of options, including the same freedom to select and independently view and follow any of the programming for which they pay. Instead, many find that ‘the current amount of available audio-described content [is] significantly below demand’ and indicate that they have difficulty finding programs with video description.''}
\end{quote}
As a result, these policies and guidelines have required broadcast, streaming, and movie theater providers to offer alternative media formats for people with cognitive or physical impairments and undeniably show an effect as more AD than ever before is indeed, available. For instance, according to the ACB AD Project `master list' \cite{ADP_AD_master} of audio-described media,
the number of unique movies and shows has increased from 1442 titles on Aug 27, 2017 to 1992 titles in June 9, 2018 -- an impressive 38\% increase in less than 12 months. 

Albeit statistics and estimates of the  aggregate of blind or visually impaired people, which are readily available on a national or global level (e.g \cite{Congdon2004, WorldHealthOrganization.2013, WHO2018} ), providing such broad glimpses of a population, rarely do those accountability efforts go into detail, especially about media use. Therefore, this paper intends to lead an empirical turn in the field of AD research and  endeavors to provide a detailed account about these audiences and their media uses, embedded in the larger framework of equal opportunity and media accessibility in the US. 

\section{Method and Data Collection}
\paragraph{Research Context and Aim}
As outlined in the prior section, policy changes and legislation implementation in the US have clearly taken the right direction to provide equally, access and opportunity for blind and visually impaired audiences. In addition, data on the ever-growing populace of blind and visually impaired people is abundant and, as outlined earlier, as a result of these policies, the media industry has been forced to increase the hours of AD on TV and expanded the selection of audio-described movies and shows in the cinema, on streaming services, and DVDs. 

However, the critical, missing piece in this relationship is rigor empirical data which provides the unfiltered account from the primary target audience of these accessibility initiatives, in the case of AD, the visually impaired and blind communities in the US. We therefore, through a national survey, investigate the perception and manifestations of those policy effects on the perceived availability, awareness, accessibility, and quality of AD in order to gather primary quantitative and qualitative data with the intention to address this contemporary, academic void. 

Our main research question considers specifically -- from the perspective of the target audience -- the current state of AD availability, accessibility, and quality...
    \begin{enumerate}
    \small
    \setlength{\itemsep}{1pt}
    \setlength{\parskip}{1pt}
    \setlength{\parsep}{1pt}
        \item ...in broadcast media, such as cable TV;
        \item ...in streaming services, such as Netflix or Amazon Prime;
        \item ...in digitally stored media, such as DVDs;
        \item ...in movie theaters and cinemas.
\end{enumerate}
To answer these research questions, this paper presents an analysis of survey responses from 483 participants, which was curated and disseminated through the ACB to its 10000 members, made available to the authors through a cooperative effort to more widely share these results. A limited overview of the survey and its results was made available on the official ACB website in late 2016 \cite{ACB_prelim}; however this paper presents the first comprehensive picture of the detailed results in the context of an academic venue. 

This survey represents as well a cornerstone of the understandings of the user base leading the design and development of `The UniDescription Project' \cite{UniD_About_US}, an AD initiative spearheaded by the University of Hawai`i at M{\=a}noa School of Communications and Center for Disability Studies, in cooperation with the US National Park Service, the ACB, and Google.

\paragraph{Survey Design and Dissemination}
The survey was developed in early 2016 and administered via SurveyMonkey, a cloud-based questionnaire and data analysis tool, from June to August 2016. The ACB promoted the survey  through digital marketing, such as social media outreach and electronic mailing list software applications (LISTSERVs) in addition to an on-site promotion in the context of the annual 2016 ACB Convention, held in Minneapolis, MN. The majority of the participants filled out the survey by themselves online, however, in cases where participants did not feel comfortable entering the answers themselves, assistance was provided by the ACB, for example during the convention or over the phone. 

The sampling method was one of convenience, in terms of which members responded to the email prompt, but multiple efforts were made to encourage and allow ACB's full membership to participate. The ACB has approximately 10000 members, with 483 respondents to our survey, we have roughly drawn a 5\% opportunity sample from the theoretical target population in this study. We further assured survey accessibility through piloting it with a small group of ACB members before full release and followed SurveyMonkey's `Best Practices' for accessible surveys \cite{surveymoneky2018}, which conform to Section 508 \cite{Rehab_act_1973_508} and the Web Content Accessibility Guidelines (WCAG) Standards. The goal of the survey was to assess the usage, opportunities, and challenges of AD in i) traditional broadcast media, such as cable TV, ii) streaming services, for example Netflix,  iii) DVDs, and iv) physical movie theaters.

\paragraph{Survey Questions and Structure}
The survey consisted of 31 questions, statements and optional comment fields to assess the awareness, independent usage  and satisfaction  with  AD  in  general,  plus  in  TV,  mobile streaming,  the  movie  theater  and  DVDs. It also assessed key demographics of the respondents (level of vision impairment, age, place of residence). Participants could choose to skip questions or statements. 

For the following analysis, the remainder of this paper will reference below survey items -- as numbered -- when presenting selected results.

\begin{enumerate}
    \setlength{\itemsep}{0pt}
    \setlength{\parskip}{0pt}
    \setlength{\parsep}{0pt}   
   \item \textbf{Key Demographics}
   \begin{enumerate}
   \small
       \item I am... \textit{(Responses: blind, visually impaired, sighted)}
       \item What is your age?: \textit{(Responses: $\leq$18, 18-25, 26-34, 35-49, 50-64, $\geq$65, No Answer)}
        \item In what state or US territory do you live? \textit{(Responses: Any US States and Territory)}
   \end{enumerate}
   \item \textbf{General Audio Description Experience}
   \begin{enumerate}
   \small
        \item Have you used audio description?	
        \item If any, please share with us the reasons why you do not use audio description?	
    \end{enumerate}
    \item \textbf{Broadcast and Cable TV}
    \begin{enumerate}
    \small
        \item Do you currently access audio description for broadcast or cable television?
        \item Do you know where to find out what content is being described on broadcast or cable television?
        \item Can you independently activate audio description on your television or through your set top box?
        \item The amount of audio-described content on broadcast or cable television meets my needs (1=strongly disagree, ..., 7=strongly agree).
        \item Approximately how many hours per week do you access audio-described content on broadcast or cable television?
        \item Please provide any additional comments on your experience accessing audio-described content on broadcast or cable television.
    \end{enumerate}
    \item \textbf{Mobile Apps and Streaming Services}
    \begin{enumerate}
    \small
        \item Do you currently access audio description for mobile apps and streaming services?
        \item Do you know where to find out what content is being audio-described on mobile apps and streaming services?
        \item Can you independently activate audio description on the mobile apps and streaming services you use?
        \item Which of the following do you use to access audio-described content on mobile apps or streaming services? 
        \item The amount of audio-described content on broadcast or cable television meets my needs  (1=strongly disagree, ..., 7=strongly agree).
        \item Approximately how many hours per week do you access audio-described content on mobile apps or streaming services?
        \item Please provide any additional comments on your experience accessing audio-described content on mobile apps and streaming services.

     \end{enumerate}
    \item \textbf{Movie Theater}
    \begin{enumerate}
    \small
       \item Do you currently access audio description for movies in the theater?
        \item Do you know where to find out what audio-described titles are available at your movie theater of choice?
        \item Does your movie theatre of choice have the equipment needed to access audio description?
        \item Is the staff knowledgeable about the assistance and equipment that is needed to access audio description?
        \item The equipment I receive always works properly (1=strongly disagree, ..., 7=strongly agree).
        \item Approximately, how often do you visit the theater to watch audio-described movies per year?
        \item Please provide any additional comments on your experience accessing audio description at the movie theater.
     \end{enumerate}
     \item \textbf{DVDs}
    \begin{enumerate}
    \small
        \item Do you currently access audio description for DVDs?
        \item Do you know how to identify which DVD titles contain audio description?
        \item Can you independently access the audio description track contained on your DVD?
        \item I am satisfied with the amount of DVDs available with audio description (1=strongly disagree, ..., 7=strongly agree).
        \item Approximately, how many DVDs with audio description do you watch each year?
        \item Please provide any additional comments on your experience accessing audio description on DVDs.
    \end{enumerate}
   \end{enumerate}


\section{Results}
\begin{figure}[!ht]
\centering
\fbox{\includegraphics[width=1\linewidth]{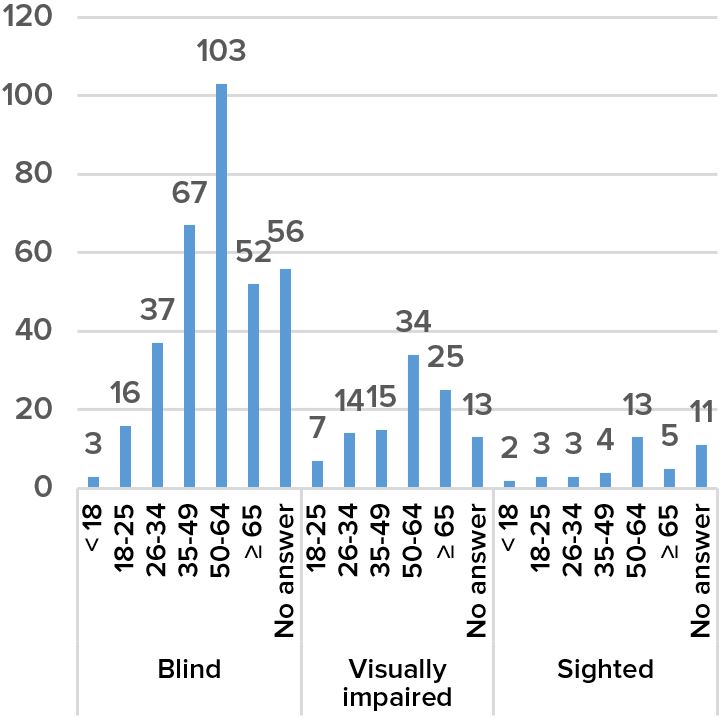}}
\caption{Question 1(a) in relation to question 1(b): Level of vision impairment and age-range.}
\label{fig:Survey_1_Age}       
\end{figure}

\paragraph{Key Demographics and General AD Experience}
In total, our survey received 483 responses. Of those who responded to question 1(a), 334 self-classified as `blind', 108 as `visually impaired', and 41 as `sighted'. The majority of our respondents were between 50 and 64 years old (Table \ref{tab:S1_Age}) and lived in either, the US East Coast or California, as indicated in Table \ref{tab:S1_Res}. Figure \ref{fig:Survey_1_Age} shows the level of visual impairment in relationship to the indicated age. 

The survey did not ask participants questions regarding gender or ethnicity, a limitation we address later on as part of the discussion. 
Out of 483 respondents, more than 90\% (447) responded yes to question 2(a), while 36 respondents answered no. 29 participants left a comment for  open-ended question 2(b). The comments' overall theme was that AD is not used due to two main factors: First, there is a general lack of awareness that AD is available in a variety of media and second, that AD is not accessible when available. We provide selected comments below:
\begin{quote} 
\small
\textit{``Because it is complicated to use.''}
\newline
\newline
\textit{``Not aware of what it is.''}
\newline
\newline
\textit{``I do not know how to use it.''}
\end{quote}

\begin{table}[!ht]
\caption{Key demographics of sample.}
\centering
\begin{subtable}{0.5\linewidth}
\caption{Question 1(b): Age-groups of participants.}
        \label{tab:S1_Age}
\centering
\begin{tabular}{|c|c|}
\hline
\textbf{Age} & \textbf{Responses} \\ \hline
$<$ 18 & 5 \\ \hline
18-25 & 26 \\ \hline
26-34 & 54 \\ \hline
35-49 & 86 \\ \hline
50-64 & 150 \\ \hline
$\geq$ 65 & 82 \\ \hline
No Answer & 80 \\ \hline
\textbf{Total} & \textbf{483} \\ \hline
\end{tabular}
\vspace{0.5cm}

\end{subtable}%


 \begin{subtable}{1\linewidth}
    \caption{Question 1(c): States indicated as place of residence in the survey (Table only shows answers with responses n$\geq$ 9).}
    \small
    \centering
\label{tab:S1_Res}
\begin{tabular}{|r|l|r|l|}
\hline
\textbf{Residence} & \textbf{Responses} & \textbf{Residence} & \textbf{Responses} \\ \hline
California & 47 & Arizona & 13 \\ \hline
Massachusetts & 42 & Maryland & 13 \\ \hline
New York & 26 & Pennsylvania & 12 \\ \hline
Florida & 25 & Minnesota & 12 \\ \hline
Texas & 20 & Michigan & 11 \\ \hline
Virginia & 19 & Missouri & 10 \\ \hline
Ohio & 15 & Washington & 9 \\ \hline
\end{tabular}
\end{subtable}
\end{table}

\begin{figure*}[ht!]
        \centering
        \begin{subfigure}[b]{0.481\textwidth}
            \centering
            \fbox{\includegraphics[width=\textwidth]{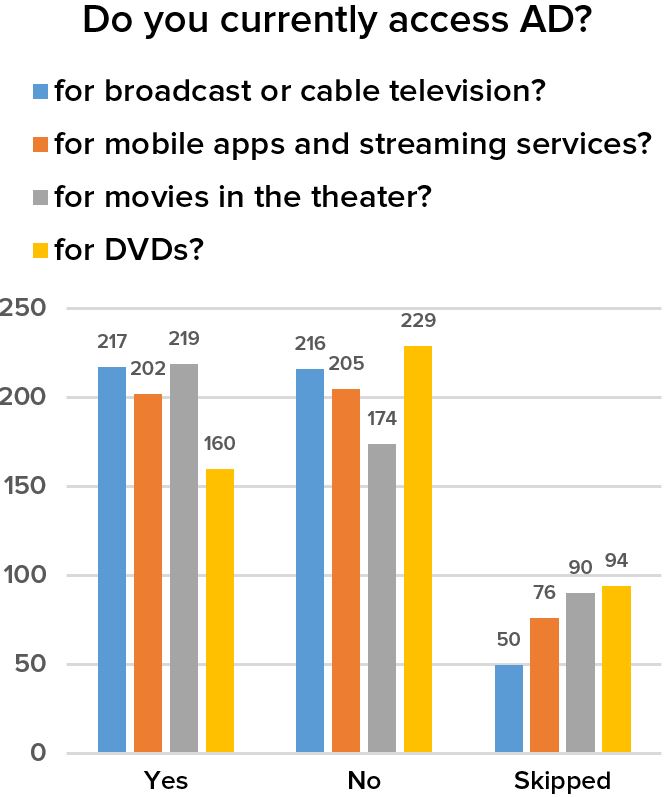}}
            \caption[]%
            {{\small Questions 3(a), 4(a), 5(a), 6(a).}}    
            \label{fig:S1_ACCESS and Utilization}
        \end{subfigure}
        \hfill
        \begin{subfigure}[b]{0.485\textwidth}  
            \centering 
            \fbox{\includegraphics[width=\textwidth]{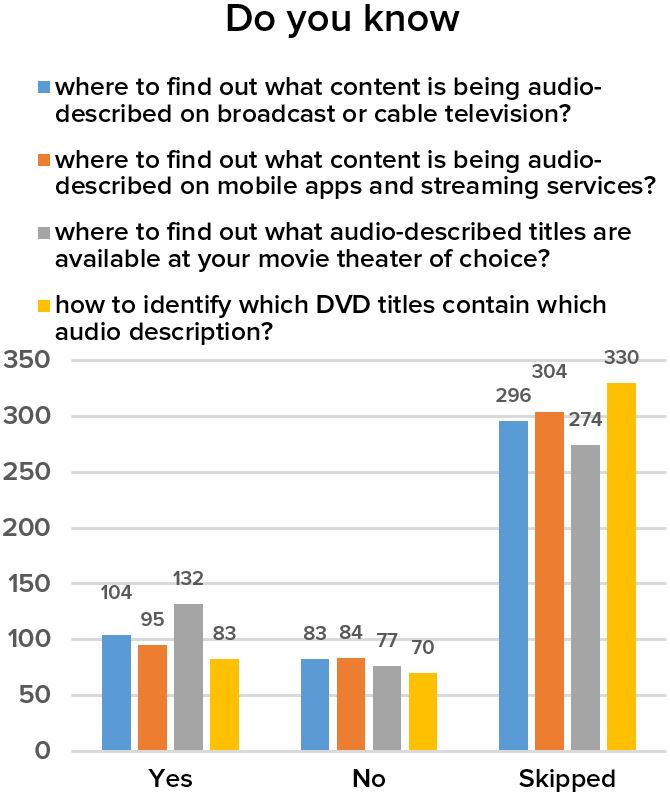}}
            \caption[]%
            {{\small Questions 3(b), 4(b), 5(b), 6(b).}}
            \label{fig:S1_AVAILABILITY}
        \end{subfigure}
        \vskip\baselineskip
        \begin{subfigure}[b]{0.4828\textwidth}   
            \centering 
            \fbox{\includegraphics[width=\textwidth]{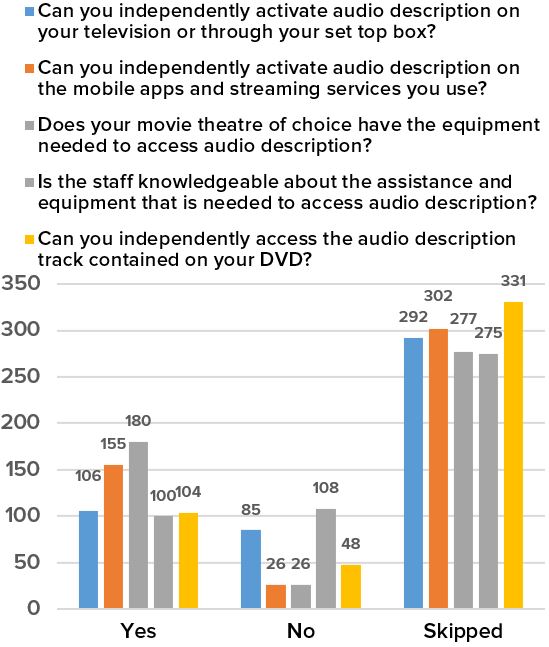}}
            \caption[]%
            {{\small Questions 3(c), 4(c), 5(c), 5(d), 6(c).}}
            \label{fig:S1_INDEPENDENCE}
        \end{subfigure}
        \hfill
        \begin{subfigure}[b]{0.485\textwidth}   
            \centering 
            \fbox{\includegraphics[width=\textwidth]{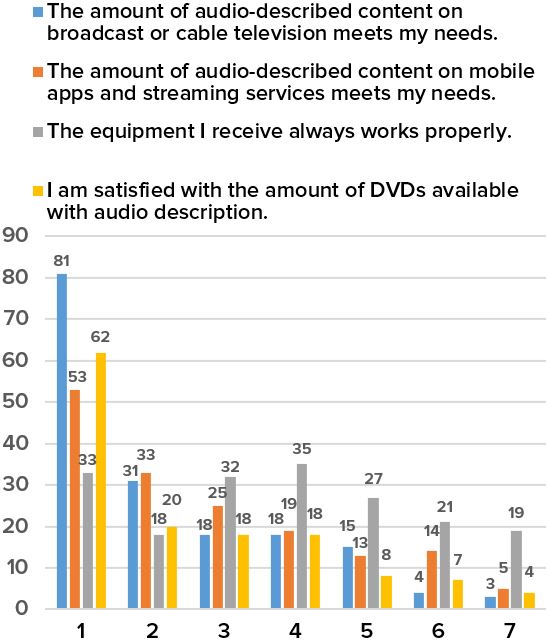}}
            \caption[]%
            {{\small Items 3(d), 4(e), 5(e), 6(d), skipped answers not shown.}}   
            \label{fig:S1_SATISFACTION}
        \end{subfigure}
        \caption{Selected survey questions and results.} 
        \label{fig:S1_Full}
    \end{figure*}

\begin{table}[!ht]
\centering
\caption{Average means for satisfaction by media format and age-groups (7-point Likert scale wherein 1=strongly disagree and 7=strongly agree). Cell colors indicate positive ($>$4.0) and negative ($\leq$4.0) average means with regards to satisfaction of AD in different media formats.}

\label{tab:means_age_media_satisfaction}
\begin{tabular}{|l|l|l|l|l|}
\hline
 & \textbf{\begin{tabular}[c]{@{}l@{}}TV\end{tabular}} & \textbf{\begin{tabular}[c]{@{}l@{}}Mobile\end{tabular}} & \textbf{\begin{tabular}[c]{@{}l@{}}Cinema\end{tabular}} & \textbf{DVDs} \\ \hline
\textbf{\begin{tabular}[c]{@{}l@{}}Average Total\end{tabular}} & \cellcolor{red!15}\textit{2.29} & \cellcolor{red!15}\textit{2.84} & \cellcolor{yellow!80}\textit{3.80} & \cellcolor{red!15}\textit{2.97} \\ \hline
\multicolumn{5}{|l|}{\textbf{Satisfaction by age group and medium formats}} \\ \hline
\textbf{$<$ 18} & \cellcolor{red!15}2 & \cellcolor{red!15}2.33 & \cellcolor{yellow!80}3.00 & \cellcolor{red!15}2.67 \\ \hline
\textbf{18-25} & \cellcolor{red!15}2.17 & \cellcolor{green!25}4.12 & \cellcolor{green!25}4.50 & \cellcolor{green!50}5.50 \\ \hline
\textbf{26-34} & \cellcolor{yellow!80}3.14 & \cellcolor{yellow!80}3.24 & \cellcolor{green!25}4.19 & \cellcolor{yellow!80}3.15 \\ \hline
\textbf{35-49} & \cellcolor{red!15}2.06 & \cellcolor{red!15}2.72 & \cellcolor{yellow!80}3.66 & \cellcolor{red!15}2.00 \\ \hline
\textbf{50-64} & \cellcolor{red!15}2.27 & \cellcolor{red!15}2.48 & \cellcolor{yellow!80}3.57 & \cellcolor{red!15}2.20 \\ \hline
\textbf{$\geq$ 65} & \cellcolor{red!15}2.06 & \cellcolor{red!15}2.16 & \cellcolor{yellow!80}3.87 & \cellcolor{red!15}2.32 \\ \hline
\end{tabular}
\end{table}


\begin{figure}[!ht]
\centering
\fbox{\includegraphics[width=1\linewidth]{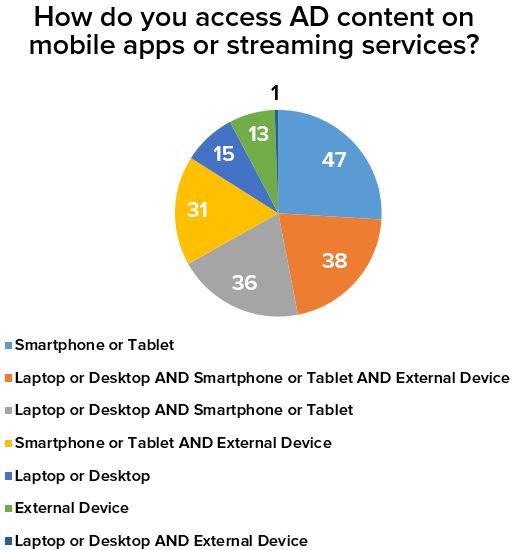}}
\caption{Question 4(d), skipped answers not shown.}
\label{fig:devices_mobile_access}       
\end{figure}

\subsection{AD in Broadcast and Cable TV}
\paragraph{Overview}The blue bars in Figure \ref{fig:S1_Full} show the responses to questions 3(a), 3(b), 3(c) and statement 3(d). Overall, the results show a balance with regards to the access and utilization of AD (blue bars, Figure \ref{fig:S1_ACCESS and Utilization}), the awareness and discovery of AD (blue bars, Figure \ref{fig:S1_AVAILABILITY}) as well as the level of independence when, for instance, starting AD on broadcast TV (blue bars, Figure \ref{fig:S1_INDEPENDENCE}). However, the responses to statement 3(d) indicate a clear dissatisfaction with the majority of respondents indicating to `strongly disagree' that AD on TV or cable meets their needs (81 out of 170 responses, blue bars, Figure \ref{fig:S1_SATISFACTION}; average mean of 2.29, Table \ref{tab:means_age_media_satisfaction}). 

From 190 responses to question 3(e), 49 participants responded to access AD less than 1 hour per week, respectively 85 for 1-3 hours, 34 for 4-6 hours, 7 for 7-9 hours and 15 responded to access AD on TV more than 9 hours (293 respondents did skip the question).

\paragraph{Qualitative comments}Participants left 121 individual comments with regards to survey item 3(f). Among those, one blind participant, older than 65 years, mentioned that: 
\begin{quote}
\small
\textit{``[...] first it is surprising the number of shows that are not audio-described for the blind, when just about everything seems to be closed captioned for the hearing impaired. Second, I have found that if I do not see a show when it is aired on cable with audio description - I miss out.  The reason is because when I try to view the show `on demand' with cable, the audio descriptive service does not work.  Third, just would like all to know that if a show has audio description, I am more likely to watch it even though I might prefer another show.''}
\end{quote}
Another participant (50-64 years old, blind) stated that he needs support from his sighted spouse to access broadcast AD: 
\begin{quote}
\small
    \textit{``Not being able to access descriptions independently is a major drawback. My sighted wife must use the visual-only onscreen menus each time and then must return the box to regular. If she does not, then some of the channels are in Spanish.''}
\end{quote}
Finding out what TV programs are audio-described seems to be a challenge, as one participant states it is:
\begin{quote}
\small
    \textit{``difficult, if not impossible, to find out what programs are AD. Finding out independently is impossible, even with a `smart' TV.''}
\end{quote} 
Other participants mentioned pros and cons of specific broadcast providers, such as the quality of customer service and ease-of-use of the provided equipment, with Comcast and Time Warner Cable as positive exemplars. We provide selected comments below:
\begin{quote}
\small
\textit{``Little to no descriptive audio....very poor selection for Verizon  fios...with fios the descriptive movies offered were only available at full price. 19.99 to watch a movie once?'' 
\newline
\newline
``My cable provider, Time Warner, passes through description on some cable channels, but not on most over-the-air channels.''
\newline
\newline
``Cable Vision is  my cable provider. I have found that when calling to ask about audio description, the people answering the phone have no idea what I am talking about.''
\newline
\newline
``I am using a bad system provided by Cox Cable.  The audio description is not always there, even when the menu says that it is available.''}
\end{quote}
Further comments stated that the hardware provided, for example the variety of set-top boxes and remote controls, is unusable by visually impaired or blind people. Participants also mentioned a `conflict of interest' of AD and Spanish on the Secondary Audio Channel (SAP). Another topic that arose was that visually impaired and blind people would de facto consume more content if it would be i) audio-described and ii) easier to access, for example, by being able to filter, browse or even permanently activate audio-described-only content on certain channels or services. 
\subsection{AD in Mobile Apps and Streaming Services}
\paragraph{Overview}The  orange  bars  in  Figure \ref{fig:S1_Full} show the responses to questions 4(a), 4(b), 4(c) and statement 4(e). The results show well-balanced answers of participants with regards to the access and utilization of AD (orange bars, Figure \ref{fig:S1_ACCESS and Utilization}) as well as the awareness and discovery of AD (orange  bars,  Figure \ref{fig:S1_AVAILABILITY}) in mobile apps and streaming services. With regard to the independent usage of AD, question 4(c), the results show a clear result: Almost six times more respondents (155 yes versus 26 no answers) indicated that they can independently activate AD in the apps and services they use (orange bars, Figure \ref{fig:S1_INDEPENDENCE}). Furthermore, Figure \ref{fig:devices_mobile_access} gives an account of the devices participants used to access AD in mobile contexts, for instance streaming services (302 participants skipped this question).  

The perceived satisfaction of our respondents with regards to the AD content on mobile apps and streaming services, statement 4(e), is overall perceived as unsatisfactory: The majority of participants indicated to `strongly disagree' as in having their needs met (orange bars, Figure \ref{fig:S1_SATISFACTION}; average mean of 2.84, Table \ref{tab:means_age_media_satisfaction}), overall indicating a clear demand for a greater variety of AD content across streaming services and mobile apps. 

From 178 responses to question 4(f), 34 participants responded to access AD less than 1 hour per week, respectively 74 for 1-3 hours, 36 for 4-6 hours, 19 for 7-9 hours and 15 responded to access AD on mobile apps or streaming services more than 9 hours per week (305 respondents skipped this question).

\begin{figure}[!ht]
\centering
\fbox{\includegraphics[width=1\linewidth]{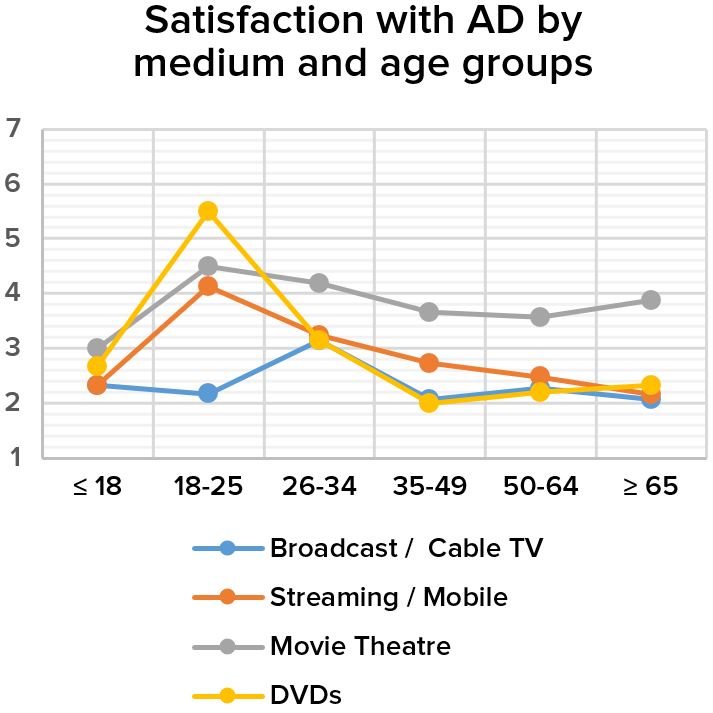}}
\caption{Line chart of satisfaction with AD by age groups and the medium format.}
\label{fig:Likert Items}       
\end{figure}

\paragraph{Qualitative comments} We received 102 qualitative comments with  regards to survey item 4(g) and will present a fraction of these herein. While comments for different streaming services were both, positive and negative, participants clearly expect more content:
\begin{quote}
\small
\textit{``I have used Netflix, Apple TV, and DVDs. I think there should be more on Netflix. And on my computer I would use it more if there's more access to more audio-described movies.''}
\newline
\newline
\textit{``My issue with accessing audio-described content is that not enough shows or movies have described movies. For example I use the HBO Go app and none of their content is audio-described. Also when I want to watch Netflix on TV, there is no way for me to access audio-described content.''}
\end{quote}
Streaming services, which are longer in the market (e.g. Netflix 1999; ITunes 2005) than others (e.g. Amazon 2006, Hulu 2007), seem to offer more audio-described content than newer providers:
\begin{quote}
\small
\textit{``Netflix and iTunes are the market leaders in this area, and the service must expand to other providers.
When will Hulu provide descriptive audio at the same level of Netflix?
Really hope Amazon video starts using audio description.''}
\end{quote}
AD quality matters too, as one participant stated in the case of Netflix, as \textit{``quality ranges from excellent for House of Cards or Grace and Frankie, to nonexistent in  The Office.''} AD can as well can cause conflict with closed captioning as one participant pointed out: 
\begin{quote}
    \small 
    \textit{``When audio description is not present I cannot read subtitles when languages, other than English, are spoken.''}
\end{quote}
Moreover,  participants mentioned the costs for individual rentals as \textit{``prohibitive''} as well as a lack of a general repository and dedicated search\footnote{Netflix, ITunes and Amazon Prime Video have by now a dedicated AD section in their services, Hulu was sued in November 2017 for not offering any AD by the ACB: \url{http://acb.org/hulu-lawsuit}} for audio-described media embedded in these services, such as \cite{ADP_AD_master}. Still, users stated that progress has been made and, for example Netflix, does  \textit{``a good job''}, while others experience the market leader as \textit{``cloogy''}.
\subsection{AD in Movie Theaters}
\paragraph{Overview}The grey bars in  Figure \ref{fig:S1_Full} show the responses to questions 5(a), 5(b), 5(c), 5(d) and  statement 5(e). The results show a slight majority (219 yes, 174 no) of respondents, who indicate to access and utilize AD in movie theaters (grey bars, Figure \ref{fig:S1_ACCESS and Utilization}). Similarly, a slight majority of respondents (132 versus 77) indicates to be aware of where -- and how -- to find information about AD at theaters (grey  bars, Figure \ref{fig:S1_AVAILABILITY}). 

With regards to questions 5(c) and 5(d), see s in Figure \ref{fig:S1_INDEPENDENCE}, the following patterns can be observed: Responses regarding the staff expertise on AD in movie theaters, respectively the quality and usability of the AD equipment in the theaters, do indicate that the majority of theaters do provide access to AD (180 yes versus 108 no answers, grey bars, Figure \ref{fig:S1_INDEPENDENCE}). However, the respondents are on the fence with regard to the assistance, knowledge and support of cinema staff (100 yes versus 108 no, grey bars, Figure \ref{fig:S1_INDEPENDENCE}). 

The perceived satisfaction of our respondents with regards to the AD equipment available for use in the movie theaters they visit, statement 5(e), is overall perceived as slightly unsatisfactory (grey bars bars, Figure \ref{fig:S1_SATISFACTION}; average mean of 3.80, Table \ref{tab:means_age_media_satisfaction}). From 212 responses to question 5(f), 127 participants indicated to visit the movie theater of their choice 1-4 times per year, respectively 50 for 5-8 times, 21 for 9-12 times and 14 participants, who indicated to visit a cinema more than 12 times a year (271 respondents skipped this question).
\paragraph{Qualitative comments}Of 121 qualitative comments left for survey item 5(g), the emerging opinion of our participants is that staff knowledge and AD equipment in cinemas varies vastly, described as \textit{``having  improved dramatically over the past couple years''} by one participant, \textit{``spotty at best''} by another and that  \textit{``staff needs more training ''} by a third participant. In addition, participants mentioned that AD devices are handed out on request to a blind patron, but are in a non-ready state: 
\begin{quote}
\small
    \textit{``Although the staff is courteous, my sighted husband often has to take my device back to customer service at the beginning of the movie because the equipment was not set right.''}
\end{quote}
Another theme we could identify is that AD devices for the blind are available, but often the staff on-site, due to a lack of training or experience, hands out wrong equipment aggravating the patrons: 
\begin{quote}
\small
\textit{
    ``While the management is informed on the AD receivers, the main staff typically have no clue. Often I end up being given a receiver for the hearing impaired because the staff do not know the difference, and they look so similar.''
    \newline
    \newline
     ``Almost every single time I attempt to receive an audio descriptive device from the movie theater I go to, the person I have to ask knows absolutely nothing about it beyond how to set it up, and almost every single time,  no matter how many times I tell them that I need audio description and not assisted listening, and even when they see my white cane, I am always given the device for the deaf  first.''
    }
\end{quote}
We retrieve a range of comments on the positive and negative aspects of the cinema experience for visually impaired or blind patrons. Overall, it seems there is no standard of service and AD in cinemas is a \textit{``hit or miss''} according to one participant, while another participant aptly commented that \textit{``some chains are better than others in terms of equipment, training, et cetera.''}

\subsection{DVDs}
\paragraph{Overview}The  yellow  bars  in  Figure \ref{fig:S1_Full} show the responses to questions 6(a), 6(b), 6(c) and statement 6(d). The results show that the majority of participants does not access audio-described DVDs (229 no versus 160 yes, yellow bars, Figure \ref{fig:S1_ACCESS and Utilization}) as well as balanced responses with regards to the ability of respondents to independently discover audio-described DVD materials (83 yes versus 70 no, yellow  bars, Figure \ref{fig:S1_AVAILABILITY}). 

With regard to the independent usage of AD on DVDs, question 6(c), the results show that about two  times more respondents can independently find and activate AD in DVDs (104 yes versus 48 no, yellow bars, Figure \ref{fig:S1_INDEPENDENCE}). The satisfaction of our respondents with the amount of audio-described DVDs, statement 6(d), is overall perceived as unsatisfactory. The majority of participants indicated to `strongly disagree' as in having their needs met by the amount of available audio-described DVD titles (yellow bars, Figure \ref{fig:S1_SATISFACTION}; average mean of 2.97, Table \ref{tab:means_age_media_satisfaction}). 

From 153 responses to question 6(e), 24 participants did indicate to watch an audio-described DVD less than two times a year, respectively 44 for 2-4 times, 39 for 5-7 times, 17 for 8-10 times and 29 responded to access DVDs with AD more than 10 times per year (330 respondents skipped this question).

\paragraph{Qualitative comments}Of 82 qualitative comments left for survey item 6(f), most of our participants mentioned that the selection of audio-described DVDs is too marginal and does not compare to what is offered to the general public:
\begin{quote}
\small
    \textit{``I wish more were described, especially DVDs of classic shows.''}
    \newline
    \newline
    \textit{``Until all DVDs are audio-described, it will never be `enough'.''}
\end{quote}
Even in cases where DVDs with AD are available, a blind person still has to  be able to first, identify the DVD, insert the DVD into the playback device, select the AD track in the DVD player and start the movie using the interface and remote control at hand -- all to be eventually able to access the audio-described content. 

Succeeding in this process is voiced as a clear issue and serious obstacle by our participants  many times, who stated to not be able to overcome independently. As one participant points out, it \textit{``is silly to provide an audio track for blind people, but provide no way to access it without sighted help.''} Mundane accessibility and usability challenges, such as identifying an audio-described DVD, using a playback software or, mastering a specific device can create crucial problems and challenges, as enunciated further in below quotes:
\begin{quote}
    \small
    \textit{``I rented 2 DVDs from my local public library [...]. A tutor who works with me had to show me how to access the audio descriptions using my MacBook's built-in DVD software.''}
   \newline
   \newline
   \textit{``[...] the biggest complaint is the inaccessible menu on DVDs. It gets very frustrating when a blind person wants to watch a DVD, but it is too much trouble, because the menu is so inaccessible.''}
   \newline
    \newline
\textit{``DVD players rarely are accessible, DVD packaging is not accessible, and menus to select description rarely are accessible.''}
 \newline
    \newline
   \textit{``Instead of having to look it up online as to whether DVDs are audio-described, they need to be labeled in braille and large print.''}
\end{quote}

\section{Discussion}
This study presents a novel, empirical account on AD in traditional and mobile media, in collaboration with one of the largest and oldest advocacy groups for blind and visually impaired people in the US. We believe that our data shows that blind or visually impaired folks are in general aware of AD, they further like and truly appreciate it - in cases where it is accessible, of high-quality, and available at no additional cost in comparison to `conventional', audio-visual media. This though, is at the present time not the case, as our quantitative and qualitative data indicates. 

Our aggregated quantitative data shows that half of our respondents stated that they can not activate AD on a TV or a set top box independently -- and that the large majority across all four media formats in this study described their overall experiences with AD as unsatisfactory (Figure \ref{fig:S1_Full}). The satisfaction across all medium formats and almost all age-groups is as well overall low, see Table \ref{tab:means_age_media_satisfaction} and Figure \ref{fig:Likert Items}, with a typical decline in satisfaction with an increase by age (with the group age of $\geq$ 65 years -- in the case of movie theaters -- being the single exception). 

On the positive side, Table \ref{tab:means_age_media_satisfaction} shows that the age group of 18-25-year-old participants was the most satisfied with the amount of AD on mobile apps (average mean of 4.12), physical DVDs (average mean of 5.50) and AD in cinemas (average mean of 4.50). Also, the age group of 26-34-year-old participants was the most satisfied with the amount of audio-described material on TV. However, the overwhelming majority of participants was overall unsatisfied as the AD available does not come up to expectations of the intended community.

Furthermore, our qualitative comments provide an account of a bandwidth of issues where AD does simply not work for the intended user. The comments also show that those, who can navigate audio-described materials in a specific ecosystem(s), do not feel that the amount of entertainment material available matches that of other interpretive or additional information, for instance closed-captioned, dubbed, or media, which has a SAP in a different language. The emerging opinion of the participants in our study indicates a clearly frustrating issue -- that (a high-quality) AD is, if anything, the very last feature of a movie or show to be thought of of.
\paragraph{Limitations}

Before we conclude and outline next steps of this research project, we point out some limitations we identify in our survey and method. First of all, we recognize that hard- and software systems, such as mobile apps and devices, user interfaces, remote controls and stationary equipment are all developing at an incredible pace. Our survey was conducted in late 2016 and we acknowledge that, due to technological advances in these areas, the availability, accessibility and usability of AD in the four media domains in this study might, in turn, have improved. 

Second, as with any empirical survey, respondents skipped certain answers, as shown in Figure \ref{fig:S1_Full} for instance, a common drawback in these type of studies, which is also known as `response fatigue' \cite{Choi.2005}. 

Third, the survey design did not assess socio-economic factors and variables of the respondents, such as ethnicity, gender, level of education or median income. These factors would potentially allow a more detailed, correlation analysis of variables in our sample population. While data on these factors would have benefited the survey, the very limited assessment of demographics (age, level of visual impairment and place of residence) was a deliberate choice, as mentioned in the prior point, with the goal to keep the survey feasible, targeted, and as concise as possible while keeping focus on the four media domains under scrutiny.

Fourth, one might argue that our population, the ACB membership base (and sample the we drew from that population), might be prone to selection and response bias. Case in point, participants might consciously undervalue the actual AD `satisfaction' to either, please the researchers or in the belief to alter media policies in a desired way. Response biases are well-researched issues in question- and survey-design, and administration \cite{Choi.2005}. We do however argue that particularly our sample of 483 participants (including 334 of these being blind) is a rarity in AD research and therefore, a clear strength of this study. 

Fifth, we concede that the research context focused on the US and is embedded in a specific legal and socio-cultural context. For example, we find significantly different policy frameworks and legal contexts in Europe (e.g.  \cite{European_Union_Standards}) or countries in South East Asia (e.g. \cite{Cogburn.2017}), accordingly limiting the applicability of our results to the US. 

Sixth, we recognize that `AD satisfaction', as in living up to one's needs, is a theoretical construct, which among others, relates to the concepts of AD quality, accessibility, availability and enjoyment. Satisfaction, as expressed in our survey, might have been of ambiguous nature to some of our respondents, although we found no evidence of that.

\section{Conclusions}
This research provides a rich, empirical account and opinion of AD in traditional and digital media, from the perspective of the visually impaired and blind community across America. In the US, legislation has been implemented which regulates availability and accessibility of AD for visually impaired and blind audiences (and the general public), and it is `enforced'. For instance, the  May 2018  FCC notice on a required 75\% increase of AD (video) description on the basis of the 21\textsuperscript{st} CCVAA  \cite[p.6]{FCC2018}, `reminded' broadcasters that:
\begin{quote}
\small
\textit{``the obligation to provide video description expands from 50
hours per calendar quarter to 87.5 hours per calendar quarter beginning July 1, 2018.''}    
\end{quote}
Although a fierce competition in the field of broadcast TV and streaming services exists, AD-based media accessibility, variety, quality of hardware, content and user interfaces differ greatly and seem to be neglected in the spirit of a successful implementation of inclusive and accessibility-oriented media policies. Indeed, many factors do play a role in the creation, proliferation, adoption and accessibility of AD through traditional and new media. Policy mandates do seem to create a legal framework which require media providers to change their programming accordingly, all in hopes of including more accessible materials for all audiences. 

\noindent However, this is not a panacea and simply mandating a certain amount of AD in any kind of media does not guarantee that it is usable, accessible, of good quality or satisfactory, the main conclusion in this paper. Therefore, as one participant stated that, \textit{``it should be a law that all content has AD if possible. If it has CC, it should have AD.''} is one of many requirements, but by no means a material conditional which axiomatically will result in usable, inclusive and accessible interfaces, devices, equipment, and media environments, to independently enjoy AD, the main observation uncovered in this study. 

With that being said, we do however, recognize that participants notice that AD is `moving into the right direction' and are mostly aware of the AD devices and services they are ought to be provided with. These devices and interfaces -- in theory available, but practically inaccessible -- as mentioned time and again by our participants in hundreds of comments, seem to be another critical hindering factor for visually impaired users and  blind patrons alike. 
\paragraph{Future Work}
In the context of the UniDescription project \cite{UniD_About_US}, we have integrated the audience in the design and evaluation process of our mobile apps as well as subject matter experts, who create high quality AD. Our paradigm is to create audio-described material that is not only available and accessible, but can provide an `equal experience' to a visual one, a challenging but ultimately true goal if one seeks to create real `equal opportunity' for everybody. 


In the future, we would like to conduct a content analysis of the rich, qualitative feedback we have received in this survey (465 comments total) and other, related field work for the purpose of an extended journal article on empirical AD research. In this limited format, we could only highlight a glimpse of the abundant feedback we received, which sheds light on the actual reasons why AD does not work in practice, even though it is implemented, available, and accessible in theory in a variety of media formats, as shown in this paper.






\bibliographystyle{ieeetr}
\bibliography{sample}

\begin{thebibliography}{10}

\bibitem{Braun2008}
S.~Braun, ``Audiodescription research: State of the art and beyond,'' {\em
  Translation Studies in the New Millennium}, vol.~6, pp.~14 -- 30, 2008.
\newblock \url{http://epubs.surrey.ac.uk/303022/}.

\bibitem{Braun2011}
S.~Braun, ``Creating coherence in audio description,'' {\em Meta}, vol.~56,
  no.~3, pp.~645--662, 2011.
\newblock \url{http://dx.doi.org/10.7202/1008338ar}.

\bibitem{Matamala2016}
A.~Matamala and P.~Orero, eds., {\em Researching Audio Description}.
\newblock London: Palgrave Macmillan {UK}, 2016.

\bibitem{Fryer.2016}
L.~Fryer, {\em An Introduction to audio description: A practical guide}.
\newblock Translation Practices Explained, London and New York: Routledge,
  2016.

\bibitem{FCC2018}
{Federal Communications Commission (FCC)}, ``{Twenty-First Century
  Communications and Video Accessibility Act},'' 2010.
\newblock
  \url{https://www.fcc.gov/general/twenty-first-century-communications-and-video-accessibility-act-0}.

\bibitem{NetflixAD}
Netflix, ``Audio descriptions for netflix movies and tv shows,'' 2018.
\newblock \url{https://help.netflix.com/en/node/25079}.

\bibitem{ADP_AD_master}
{Audio Description Project, American Council of the Blind}, ``{Master List of
  Audio-described Videos},'' 2018.
\newblock \url{http://www.acb.org/adp/masterad.html}.

\bibitem{Rehab_act_1973_504}
{United States Department of Labor}, ``{Section 504, Rehabilitation Act of
  1973},'' 2018.
\newblock \url{https://www.dol.gov/oasam/regs/statutes/sec504.htm}.

\bibitem{Rehab_act_1973_508}
{United States General Services Administration}, ``{IT Accessibility Laws and
  Policies - Section508.gov},'' 2018.
\newblock \url{https://www.section508.gov/manage/laws-and-policies}.

\bibitem{Disability_Act1990}
{United States Department of Labor}, ``{Americans with Disabilities Act },''
  2018.
\newblock \url{https://www.dol.gov/general/topic/disability/ada}.

\bibitem{FCC_Fact_Sheet_2017}
{Federal Communications Commission (FCC)}, ``{FCC FACT SHEET: Video Description
  Expansion},'' 2017.
\newblock \url{https://apps.fcc.gov/edocs_public/attachmatch/DOC-345472A1.pdf}.

\bibitem{Congdon2004}
N.~Congdon, B.~O'Colmain, C.~C.~W. Klaver, R.~Klein, B.~Muñoz, D.~S. Friedman,
  J.~Kempen, H.~R. Taylor, P.~Mitchell, and {Eye Diseases Prevalence Research
  Group}, ``Causes and prevalence of visual impairment among adults in the
  united states,'' {\em Archives of ophthalmology (Chicago, Ill. : 1960)},
  vol.~122, p.~477—485, April 2004.

\bibitem{WorldHealthOrganization.2013}
{World Health Organization}, {\em Universal eye health: A global action plan
  2014-2019}.
\newblock Geneva: {World Health Organization}, 2013.
\newblock \url{http://www.who.int/blindness/AP2014_19_English.pdf}.

\bibitem{WHO2018}
{World Health Organization}, ``Vision impairment and blindness,'' 2018.
\newblock \url{http://www.who.int/mediacentre/factsheets/fs282/en/}.

\bibitem{ACB_prelim}
{Anthony Stephens}, ``Acb survey finds need for increased audio description,''
  2016.
\newblock \url{http://acb.org/Audio-Description-Survey}.

\bibitem{UniD_About_US}
{UniDescription.org}, ``{UniD - About Us},'' 2018.
\newblock \url{https://www.unidescription.org/about}.

\bibitem{surveymoneky2018}
{SurveyMonkey}, ``{Web Accessibility: 508 Compliance \& WCAG2},'' 2018.
\newblock \url{https://help.surveymonkey.com/articles/en_US/kb/508-Compliance}.

\bibitem{Choi.2005}
B.~C.~K. Choi and A.~W.~P. Pak, ``A catalog of biases in questionnaires,'' {\em
  Preventing chronic disease}, vol.~2, no.~1, p.~A13, 2005.

\bibitem{European_Union_Standards}
{European Union Agency for Fundamental Rights}, ``{Accessibility standards for
  audio-visual media},'' 2018.
\newblock
  \url{http://fra.europa.eu/en/publication/2014/indicators-right-political-participation-people-disabilities/audiovisual-standards}.

\bibitem{Cogburn.2017}
D.~L. Cogburn and T.~{Kempin Reuter}, eds., {\em {Making disability rights real
  in Southeast Asia: Implementing the UN Convention on the Rights of Persons
  with Disabilities in ASEAN}}.
\newblock Lanham: {Lexington Books}, 2017.

\end{thebibliography}

\end{document}